\documentclass[twocolumn,aps,pre]{revtex4}
\usepackage{graphicx}
\usepackage{dcolumn}
\usepackage{amsmath}
\usepackage{latexsym}

\begin {document}

\title {Double Transition in a Model of Oscillating Percolation}
\author
{Sumanta Kundu$^1$, Amitava Datta$^2$, and S. S. Manna$^1$}
\affiliation
{
\begin {tabular}{c}
$^1$Satyendra Nath Bose National Centre for Basic Sciences,
Block-JD, Sector-III, Salt Lake, Kolkata-700106, India \\
$^2$School of Computer Science and Software Engineering, University of Western Australia, Perth, WA 6009, Australia
\end{tabular}
}
\begin{abstract}
      Two distinct transition points have been observed in a problem of lattice percolation studied using a 
   system of pulsating discs. Sites on a regular lattice are occupied by circular discs whose radii vary
   sinusoidally within $[0,R_0]$ starting from a random distribution of phase angles. A lattice bond is said to be connected
   when its two end discs overlap with each other. Depending on the difference of the phase angles of these
   discs a bond may be termed as dead or live. While a dead bond can never be connected, a live bond is
   connected at least once in a complete time period. Two different time scales can be associated with
   such a system, leading to two transition points. Namely, a percolation transition occurs at $R_{0c} = 
   0.908$ when a spanning cluster of connected bonds emerges in the system. Here, information propagates
   across the system instantly, i.e., with infinite speed. Secondly, there exists another transition point
   $R_0^* = 0.5907$ where the giant cluster of live bonds spans the lattice. In this case the information 
   takes finite time to propagate across the system through the dynamical evolution of finite size clusters. 
   This passage time diverges as $R_0 \to R_0^*$ from above. Both the transitions exhibit the critical behavior 
   of ordinary percolation transition. The entire scenario is robust with respect to the distribution of frequencies
   of the individual discs. This study may be relevant in the context of wireless sensor networks.
\end{abstract}

\maketitle

\section{Introduction}

      The beauty of percolation model lies in its simplicity as well as non-triviality in studying the
   order-disorder phase transition \cite {Stauffer,Grimmett,Meester}.
%     Percolation is a well known model that exhibits an order-disorder phase transition \cite
%  {Stauffer,Grimmett,Meester}. 
   A number of variants of the percolation model have been introduced in last several decades 
   \cite {Araujo,Saberi,Lee,Achlioptas,Coniglio}. The theory of percolation has been 
   successfully applied to a variety of problems such as metal-insulator transition \cite {Arcangelis}, 
   epidemic spreading in a population \cite {Grassberger, Newman}, gelation in polymers \cite {Flory}, 
   wireless communication networks \cite {Dargie,Glauche,Krause} etc. The generic feature of all percolation 
   models is the appearance of long range connectivity from the short range connectedness when the control 
   variable is tuned to the critical point \cite {Sahimi}. The critical points of the percolation models 
   are dependent on the geometry of the system, whereas their critical behavior is characterized by a 
   universal set of critical exponents \cite{Stauffer}.
    
      Wireless sensor networks (WSN) \cite{Dargie} are usually composed of sensor nodes which are deployed 
   in a regular topology in the form of a grid for collecting various environmental data, e.g., temperature 
   and humidity. Often a sensor node has a low-powered radio, limited processing and storage capabilities. 
   Hence it is important that nodes can send collected data to a base station using a multi-hop radio link 
   through the intermediate nodes in a WSN. 

      The wireless range of each node is approximately circular and a direct path is established when a node 
   becomes connected to the base station through overlapping wireless ranges of intermediate nodes. This 
   problem is similar to the percolation problem as the base station and a transmitting node becomes part of 
   a percolating cluster when a radio link is established through overlapping radio transmission ranges of 
   intermediate nodes. It is well known that the wireless ranges of low-power sensor nodes fluctuate temporally 
   due to interference and noise \cite {Sollacher,Baccour1,Baccour2}. It is important to know when such a percolating 
   path exists as a sensor node can then transmit its packets to the base station without any need for buffering 
   the packets in intermediate nodes, as each sensor node has very little buffer space and packets that cannot 
   be immediately transmitted are usually dropped. Our study of the oscillating percolation problem is a first 
   attempt in understanding percolation in the presence of such time-varying transmission ranges. Such temporal 
   variations exist in above-ground \cite {Baccour1,Baccour2}, above-ground to underground \cite {Silva} and 
   aerial-sensor networks \cite {Ahmed}. The speed of variation of these transmission ranges is usually much 
   slower compared to radio transmission speed and hence a percolating cluster persists for a long enough 
   duration for transmitting packets in a WSN.

%--------------------------------------------------------------------------
\begin{figure*}[t]
\begin{center}
\begin {tabular}{cccc}
\includegraphics[width=4.3cm]{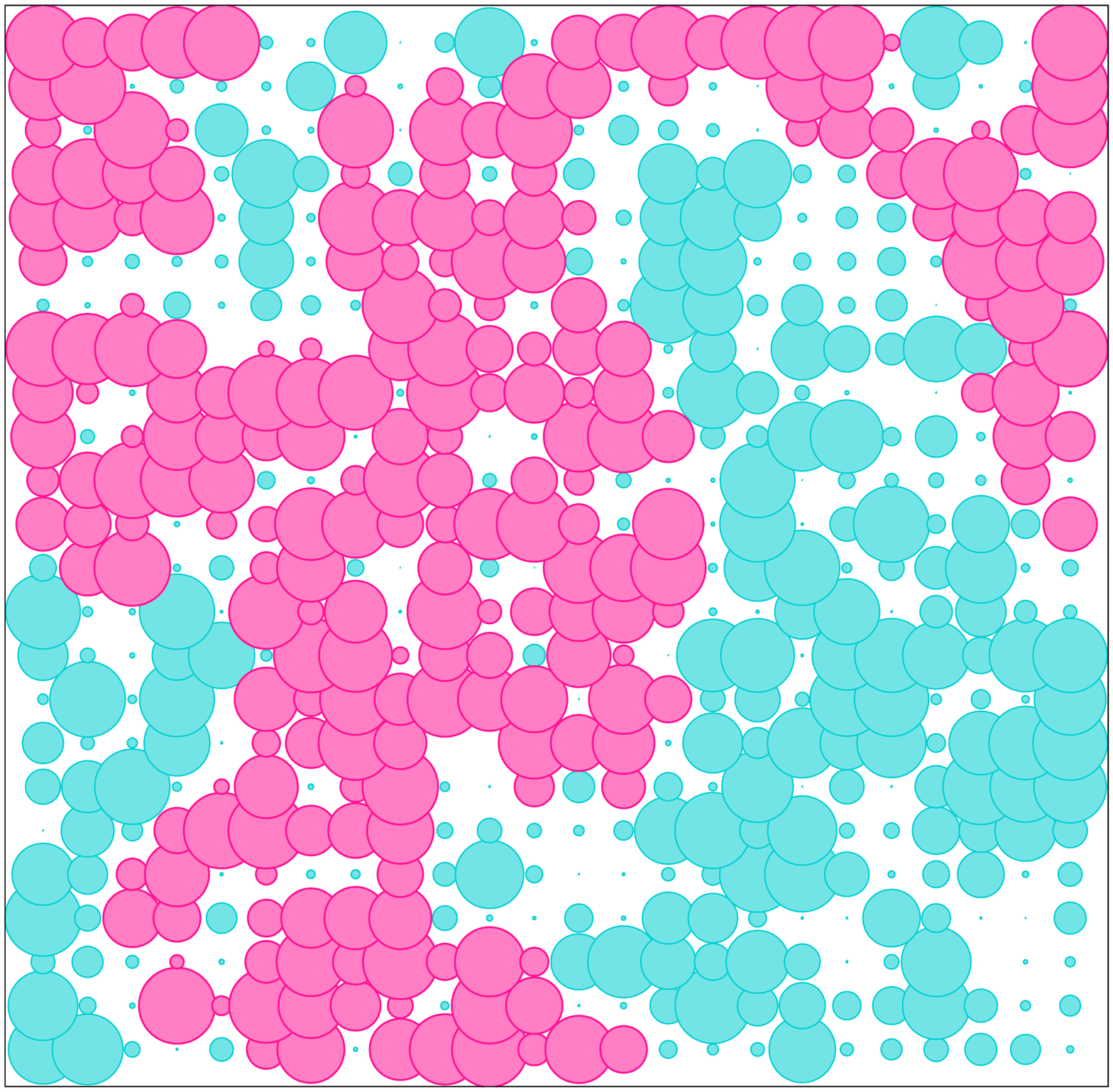} & \includegraphics[width=4.3cm]{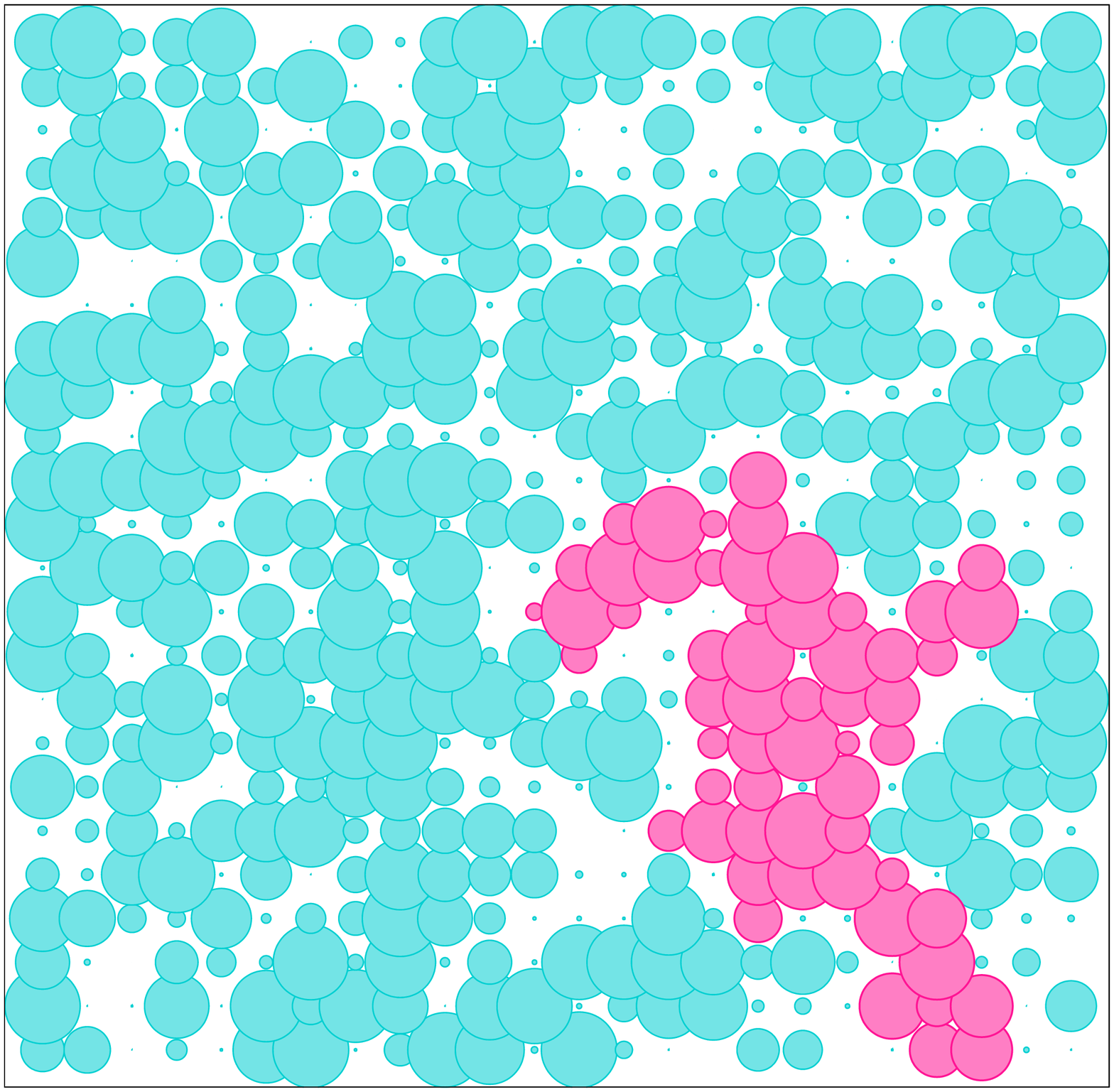} &
\includegraphics[width=4.3cm]{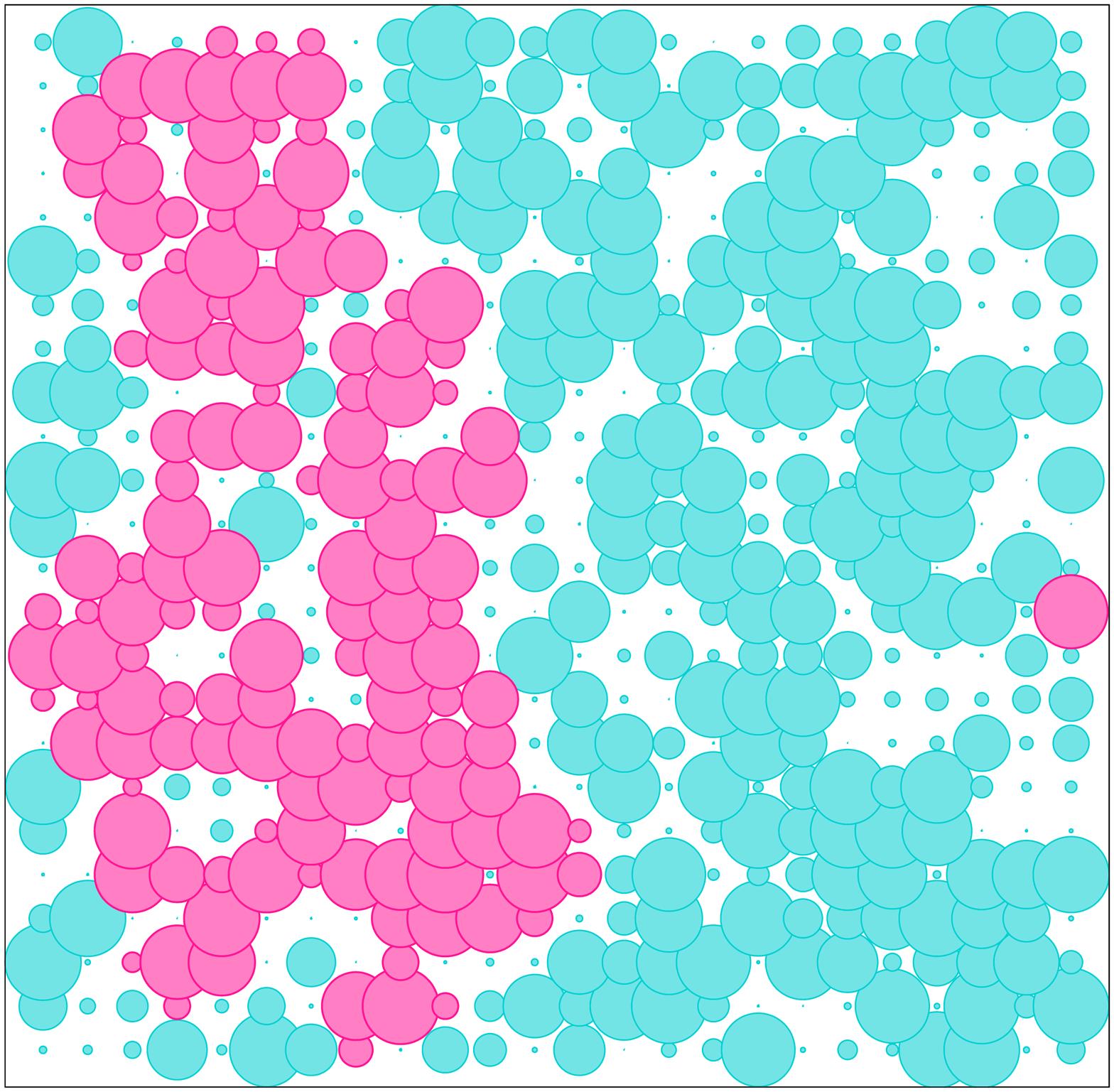} & \includegraphics[width=4.3cm]{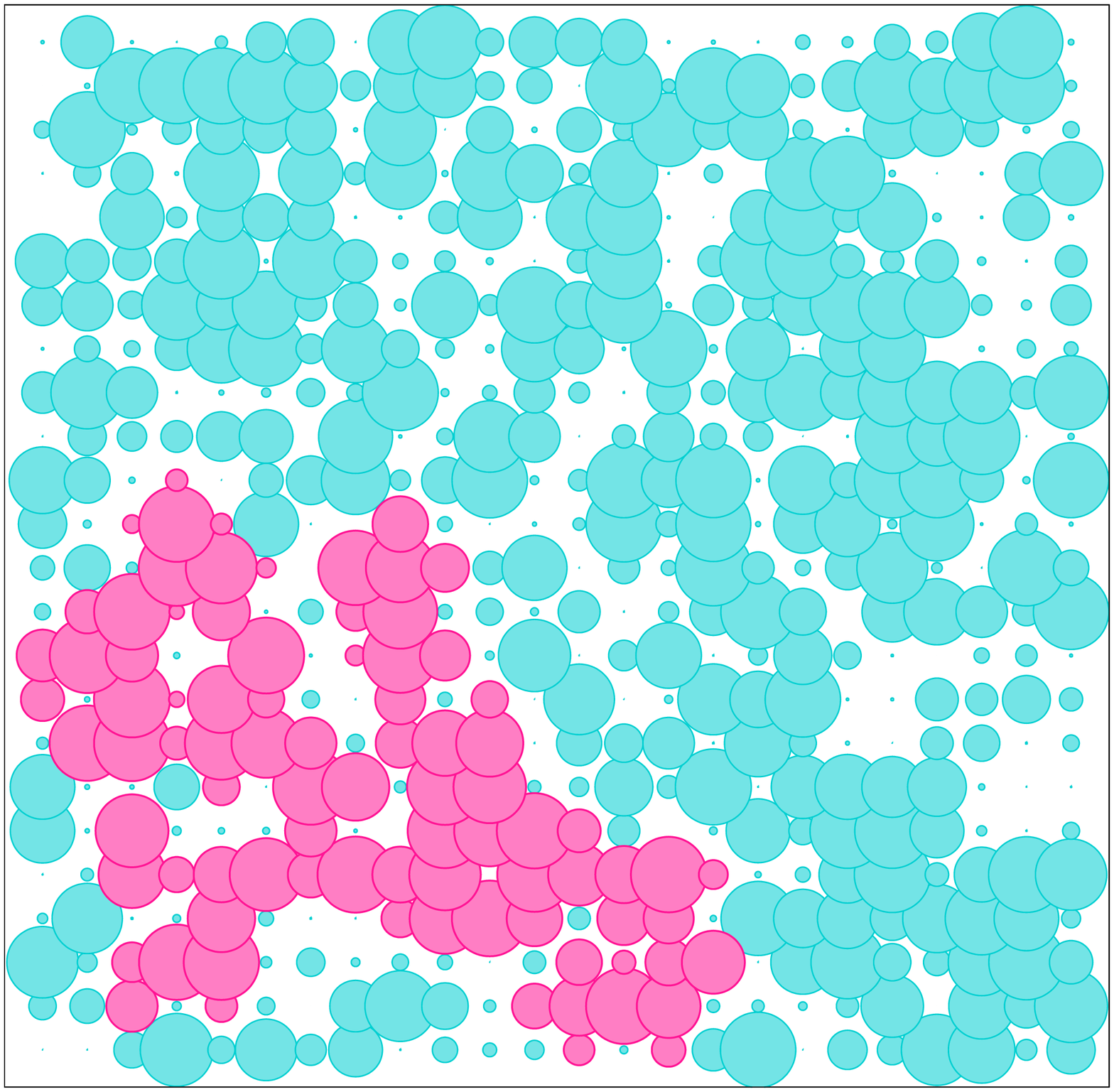}
\end {tabular}
\end{center}
\caption{Snapshots of the time dependent percolation configuration have been shown on a square lattice 
   of size $L = 24$ with periodic boundary conditions along the horizontal direction. The radii of all 
   the discs having angular frequency $\omega = 1$ pulsate with time as per Eqn.\ \ref{EQN01} and are 
   different at a given time $t$ due to the random initial phases $\{\phi\}$. For $R_0 = 0.85$, the 
   snapshots are taken at $t = 150 dt, 300 dt, 500 dt$ and $600 dt$ (from left to right), where $dt = 
   \pi/L^2$. The largest cluster painted in magenta sometimes spans the entire lattice and sometimes 
   does not.}
\label{FIG01}
\end{figure*}
%---------------------------------------------------------------------------
 
      In this paper, our objective is to model the temporal fluctuations of radio transmission ranges in 
   the WSNs using the framework of percolation theory. Sites of a square lattice are occupied by
   circular discs of time-varying radii $R(t)$ which pulsate sinusoidally, mimicking the temporal variations 
   of the radio transmission ranges of sensor nodes. Accordingly, a bond between a pair of neighboring 
   sites is considered to be connected if and only if the discs at these sites overlap. Initial assignment 
   of random phase angles of the pulsating discs makes the system heterogeneous. Therefore, the duration of time 
   that a bond remains connected depends on the phases of the two end discs and is different for different 
   bonds. The maximal value of disc radii $R_0$ is the same for all discs and is the control variable of the
   problem. In some instants of time the system may be globally connected through the spanning paths of 
   connected bonds between the opposite boundaries of the lattice. In the time averaged description, the 
   system undergoes a continuous percolation transition for $R_0 > R_{0c}$ for the infinitely large system. 
   Further, for $R_0 < R_{0c}$ when there exists no spanning path, information can still propagate across 
   the system through different finite size clusters of connected bonds which appear in different instants 
   of time, if longer propagation time is allowed. On the average this transmission time increases as $R_0$ 
   is decreased and it diverges as $R_0 \rightarrow R_0^*$ from above. In the following we present evidence 
   that the system undergoes a second percolation transition at this point. We have studied the critical 
   properties of the system around both the transition points. This study may also be relevant in the context of
   spreading of epidemic disease in a population, spreading of computer viruses through the Internet, and 
   even for rumor spreading in the social media etc. 

      The paper is organized as follows. We start by describing the model of oscillating percolation in 
   Sec.\ II. The connectivity properties of lattice bonds are investigated in Sec.\ III. The calculation of the order 
   parameter and the spanning probability is described in Sec.\ IV. In Sec.\ V, we discuss the dependence 
   of the percolation properties on the frequencies of the pulsating discs. In Sec.\ VI, we have observed
   the existence of a second percolation transition point defined in terms of two time scales for the speed
   of information propagation through the connected clusters. In Sec.\ VII, we generalize the model of 
   oscillating percolation. Finally, we summarize in Sec.\ VIII.

\section{Model}

      A circular disc of radius $R(t)$ that varies with time $t$ has been placed at every site of a square 
   lattice of size $L \times L$ with unit lattice constant. The radii of the discs pulsate periodically 
   following a sinusoidal variation as:
\begin{equation}
R(t) = (R_0/2)[\sin(\omega t + \phi) + 1]
\label{EQN01}
\end{equation}
   where, $R_0$ is the control variable that varies in the range $[0,1]$; the phase $\phi$ and the angular 
   frequency $\omega$ being two parameters. At time $t$=0, every site is assigned a disc of radius $R(0)$ 
   with a random phase angle drawn from a uniform probability distribution $p(\phi) =1/2\pi$, $0 \leq \phi 
   < 2\pi$. With this only randomness in phase angles, the radii of the discs start pulsating between $[0,R_0]$
   following Eqn.\ \ref{EQN01} in a completely deterministic fashion.

\begin{figure}[t]
\begin{center}
\includegraphics[width=\linewidth]{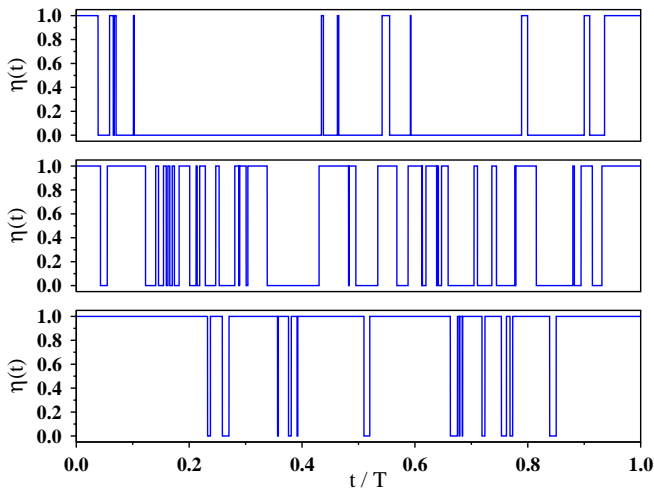}
\end{center}
\caption{For $\omega = 1$ and $L$ = 128, the phase representing variable $\eta(t)$ has been plotted 
   with $t$ during a period $T$ for $R_0$ = 0.88, 0.90 and 0.92 (from top to bottom). The value of 
   $\eta(t) = 1$ and $0$ correspond to the percolating and non-percolating phases respectively.}
\label{FIG02}
\end{figure}

      A bond between a pair of neighboring discs of radii $R_1(t)$ and $R_2(t)$ is defined to be connected 
   only when
\begin{equation}
R_1(t) +R_2(t) \geq 1,
\label{EQN02}
\end{equation}
   which is referred as the Sum Rule. The connection status of every bond over a period $T = 2\pi / \omega$ 
   would be repeated ad infinitum. A group of sites interlinked through the connected bonds forms a cluster. 
   At a particular time there are several clusters of different shapes and sizes. During the time evolution 
   sometimes the largest cluster spans the entire lattice and establishes a global connection (Fig.\ \ref{FIG01}). 
   Therefore, within one time period $T$, the system in general switches between the percolating and non-percolating 
   states. We define a flag $\eta(t)$ = 1 and 0 for the percolating and non-percolating states respectively 
   and its variation is exhibited in Fig.\ \ref{FIG02}. The average residence time in percolating state
   increases on increasing $R_0$. To estimate how much the disc configuration becomes different from its initial 
   configuration in time $t$ we define a hamming distance $\Delta(t) = \max\{|R_i(t)-R_i(0)|\}$ calculated 
   over all sites $i$ which is found to vary as $\Delta(t) = R_0 \sin(\pi t/T)$.

%--------------------------------------------------------------------------
%--------------------------------------------------------------------------- 

%--------------------------------------------------------------------------
%\begin{figure}[t]
%\begin{center}
%\begin {tabular}{c}
%\includegraphics[width=\linewidth]{Figure03.eps}
%\end {tabular}
%\end{center}
%\caption{The difference $\Delta(t)$ (defined in the text) of the time dependent disc configuration at time $t$ 
%   from the same configuration at time $t=0$ generated using random values of phases $\{\phi\}$, has been plotted 
%   against $t/T$ on a square lattice of size $L = 64$, for $R_0$ = 1.0 (black), 0.8 (red) and 0.6 (blue) using 
%   $\omega = 1$ for all the discs (arranged from top to bottom).}
%\label{FIG03}
%\end{figure}
%---------------------------------------------------------------------------

\section{Connectivity of the Bonds}

      The phase difference between the two pulsating discs at the ends of a bond has a crucial role for the connectivity 
   of the bond. For $R_0 = 1/2$, the bond is connected only at a single instant within the time period $T$ if the discs 
   are in the same phase, whereas, for $R_0 = 1$ the bond remains always connected if the discs are in the opposite phase. 
   This implies that for $1/2 < R_0 < 1$, a bond is connected within a period $T$ only when the phase difference of the two 
   end discs lies within a certain range. The maximum value of the sum $R_1(t) + R_2(t)$ must be $R_0[\cos(\Delta \phi /2) + 1] \ge 1$ 
   for the bond to be connected and
\begin{equation}
\Delta \phi = |\phi_2 - \phi_1| \leq \Delta \phi_c = 2 \cos^{-1}\Big(1/R_0 - 1\Big).
\label{EQN03}
\end{equation}
   Evidently, this range increases with increasing the value of $R_0$.
   The fraction of time over which a bond remains connected within a period $T$ is given by,
\begin{equation}
f_T(R_0,\Delta \phi) = 1/2 - (1/\pi) \sin^{-1}\Big((1/R_0 - 1)\sec(\Delta \phi /2)\Big).
\label{EQN04}
\end{equation}
   For a connected bond we must have $f_T(R_0,\Delta \phi) \geq 0$ which also gives
   Eqn.\ \ref{EQN03}. For the special case of $R_0 = 1$ and $\Delta \phi = \pi$ we get $f_T = 1$.

%--------------------------------------------------------------------------
\begin{figure}[t]
\begin{center}
\includegraphics[width=\linewidth]{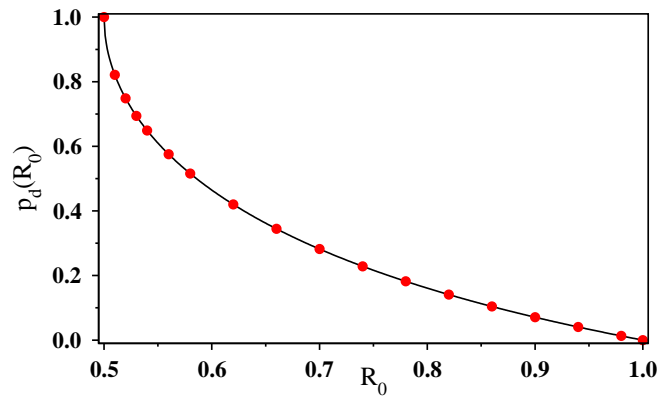}
\end{center}
\caption{Plot of the density of dead bonds $p_d(R_0)$ against $R_0$ which never get connected during the entire time evolution.
   The numerically obtained data for system size $L = 256$ (filled circles) fit very well with the 
   functional form given in Eqn.\ \ref{EQN05} (solid line).}
\label{FIG03}
\end{figure}
%---------------------------------------------------------------------------

      This implies that a bond remains unconnected forever if $\Delta \phi > \Delta \phi_c$. We call these bonds 
   as the `dead' bonds. In contrast, the remaining set of bonds dynamically changes their connectivity status 
   within a period $T$ and are referred as the `live' bonds. The densities of dead and live bonds are denoted by 
   $p_d$ and $p_l$ respectively. Expectedly, $p_d$ increases when $R_0$ is decreased from 1 and it approaches 
   unity as $R_0 \to 1/2$. Since $p(\phi)$ is uniform, the quantity $p_d(R_0)$ is calculated as
\begin{eqnarray}
p_d(R_0) &=& 1 - 2p(\phi)\Delta \phi_c \nonumber \\
            &=& 1 - (2/\pi) \cos^{-1}\Big(1/R_0 - 1\Big).
\label{EQN05}
\end{eqnarray}
   In Fig.\ \ref{FIG03}, good agreement is observed between the plots of the numerically estimated values of 
   $p_d(R_0)$ against $R_0$ and the functional form given in Eqn.\ \ref{EQN05}. 

\section{The Order Parameter and the spanning probability}

      The order parameter $\Omega(R_0,L)$ is defined as the fractional size of the largest cluster, doubly
   averaged with respect to time between 0 and $T$ and over many initial configurations ${\cal C}$ with 
   different sets of random phase angles $\{\phi_i\}$.
\begin{equation}
\Omega(R_0,L) = \langle \langle s_{max}(R_0,L) \rangle_T \rangle_{\cal C} / L^2.
\label{EQN06}
\end{equation}
   We also define $\Pi(R_0,L)$ as the spanning probability from the top to the bottom of the lattice.

      In numerical simulations time is increased in equal steps of $dt = T/(2L^2)$. Periodic
   boundary condition has been imposed along the horizontal direction, whereas the global connectivity is checked
   along the vertical direction as in cylindrical geometry. Both the order parameters $\Omega(R_0,L)$ and 
   $\Pi(R_0,L)$ are estimated for a large number of values of $1/2 < R_0 \le 1$ with a minimum increment of 
   $\Delta R_0 = 0.001$.

%--------------------------------------------------------------------------
\begin{figure}[t]
\begin{center}
\includegraphics[width=\linewidth]{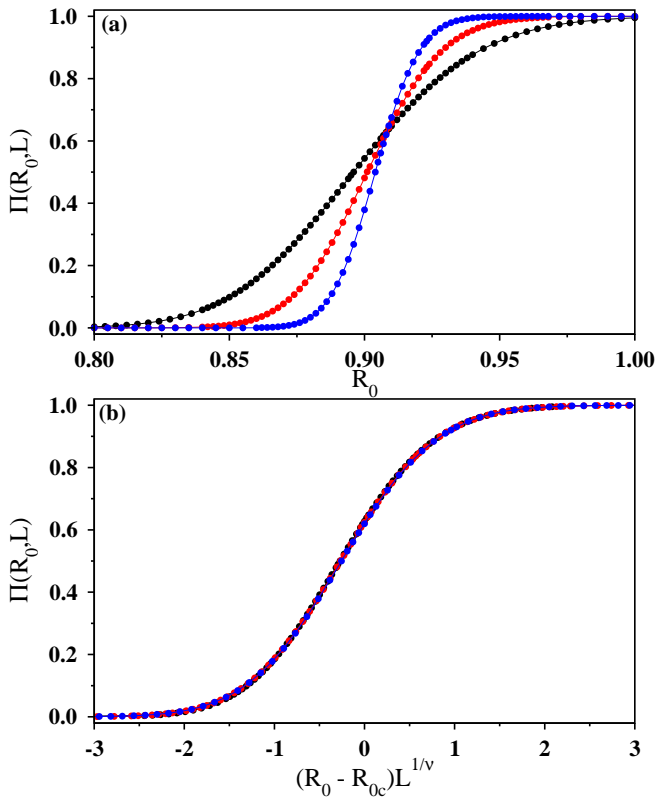}
\end{center}
\caption{For $\omega = 1$, and the system sizes $L$ = 64 (black), 128 (red), and 256 (blue) (arranged from left to right). (a) The spanning 
   probability $\Pi(R_0,L)$ has been plotted against $R_0$. (b) Finite-size scaling plot of the same data with 
   $R_{0c} = 0.908$ and $1/\nu = 0.75$ exhibits a very nice data collapse.}
\label{FIG04}
\end{figure}
%---------------------------------------------------------------------------
%--------------------------------------------------------------------------
\begin{figure}[t]
\begin{center}
\includegraphics[width=\linewidth]{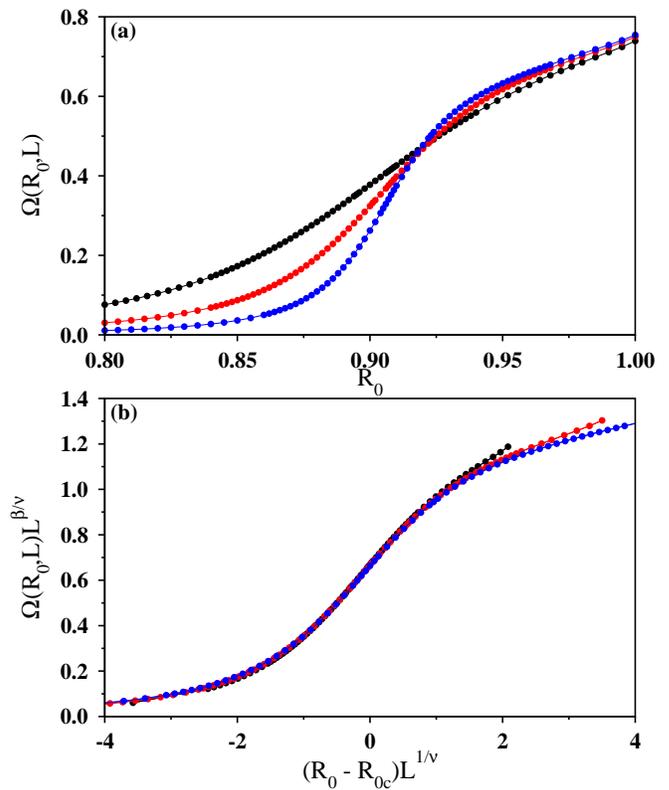}
\end{center}
\caption{For $\omega = 1$, (a) variation of the order parameter $\Omega(R_0,L)$ as defined in Eqn.\ \ref{EQN06} 
   with $R_0$ has been shown for the system sizes $L$ = 64 (black), 128 (red), and 256 (blue) (arranged from left to right); (b) finite-size 
   scaling of the same data using $R_{0c} = 0.908(5)$, $1/\nu = 0.75$ and $\beta/\nu = 0.114(5)$ exhibits an excellent 
   data collapse.}
\label{FIG05}
\end{figure}
%---------------------------------------------------------------------------

      In Fig.\ \ref{FIG04}(a), $\Pi(R_0,L)$ has been plotted against $R_0$ for three different system sizes using 
   $\omega = 1$ for all discs. These curves intersect approximately at the point $(R_{0c},\Pi(R_{0c}))$. 
   We estimate $R_{0c} \approx 0.90$ and the spanning probability $\Pi(R_{0c}) \approx 0.63$ which is quite
   consistent with the value $0.636454001$ \cite{Ziff} obtained using Cardy's formula \cite{Cardy}.
   For a more precise estimation of $R_{0c}$ we define $R_{0c}(L)$ for individual system sizes by
   $\Pi(R_{0c}(L),L) = 1/2$. The $R_{0c}(L)$ values are estimated by linear interpolation of the data in 
   Fig.\ \ref{FIG04}(a) and then extrapolated to $L \to \infty$ to obtain $R_{0c}$. Tuning the value of
   $R_{0c}$ the difference $R_{0c} - R_{0c}(L)$ has been plotted against $L^{-1/\nu}$ to obtain the best 
   value of $R_{0c} = 0.908(5)$. Here $\nu = 4/3$, the correlation length exponent of ordinary percolation.
   Further, for a finite size scaling plot $\Pi(R_0,L)$ has been plotted against $(R_0 - R_{0c})L^{1/\nu}$.
   An excellent data collapse for all three system sizes in Fig.\ \ref{FIG04}(b) indicates the finite-size 
   scaling form:
\begin{equation}
\Pi(R_0,L) \sim {\cal G} \big[(R_0 - R_{0c})L^{1/\nu}\big].
\label{EQN07}
\end{equation}
   A similar analysis has been performed for the order parameter $\Omega(R_0,L)$. Figure \ref{FIG05}(a)
   shows $\Omega(R_0,L)$ against $R_0$ plot for the same three system sizes and their finite-size scaling 
   analysis have been done in Fig.\ \ref{FIG05}(b), indicating the scaling form:
\begin{equation}
\Omega(R_0,L)L^{\beta/\nu} \sim {\cal F} \big[(R_0 - R_{0c})L^{1/\nu}\big].
\label{EQN08}
\end{equation}
   From this scaling we get $\beta/\nu = 0.114(5)$ compared to the exact values $\beta/\nu =5/48 \approx 0.1042$ 
   with $\beta = 5/36$ \cite{Stauffer} for ordinary percolation.

\section{Percolation with distributed frequencies}

\begin{figure}[t]
\begin{center}
\includegraphics[width=\linewidth]{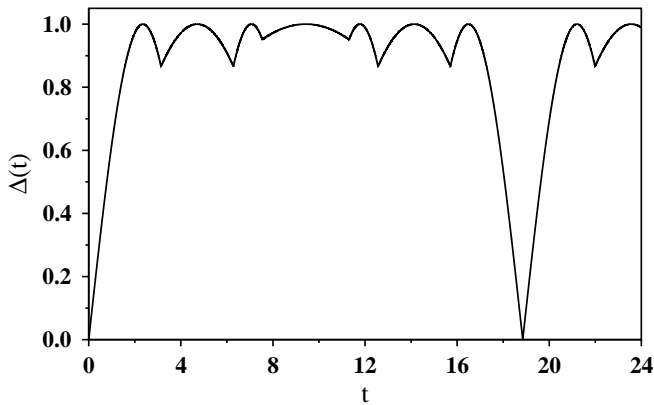}
\end{center}
\caption{Plot of $\Delta(t)$ against $t$ for $L=64$, $R_0=1$ and only for one initial configuration. The 
   system is composed of three different types of discs characterized by their own frequencies: $\omega_1 
   = 1/3, \omega_2 = 2/3$ and $\omega_3 = 4/3$. Here, we find that the minimum value of $\Delta(t) = 2.96 
   \times 10^{-4}$ is at $t = 18.85$. The numerical estimate of $T = 18.85$ matches considerably well with 
   the value of $T = 6\pi$ calculated using Eqn.\ \ref{EQN10}.}
\label{FIG06}
\end{figure}

      Now we consider the situation where each disc is randomly assigned a frequency $\omega_1$ with probability 
   $f$ and frequency $\omega_2$ with probability $1 - f$ with previously prescribed random phase angles.
   The time period $T_{(\omega_1,\omega_2)}$ has been calculated numerically for a large number of pairs of 
   angular frequencies, where the frequencies are the rational numbers. 
   Since for two rational numbers $a/b$ and $c/d$, HCF$(a/b, c/d)$ = HCF$(a,c)$ / LCM$(b,d)$, HCF and LCM being the 
   highest common factor and lowest common multiplier respectively, we conjecture the following functional
   form
\begin{equation}
T_{(\omega_1,\omega_2)} = 2\pi/\text{HCF}(\omega_1,\omega_2)
\label{EQN09}
\end{equation}   
   which is independent of $0 < f < 1$.
   A generalized form of the above expression for $T$ can further be conjectured for the mixture of $N$ distinct
   frequencies $\omega_1, \omega_2, ..., \omega_N$ as
\begin{equation}
T_{(\omega_1,\omega_2,...,\omega_N)} = 2\pi/ \text{HCF}(\omega_1,\omega_2,...,\omega_N).
\label{EQN10}
\end{equation}     
   This conjecture has been numerically verified using the mixtures up to five distinct frequencies. For example, the time 
   period is estimated using the plot of $\Delta(t)$ against $t$ in Fig.\ \ref{FIG06} for three distinct frequencies.

%--------------------------------------------------------------------------
%---------------------------------------------------------------------------

      This model is further extended by assigning a distinct frequency to each disc drawing them from a
   uniform distribution $p(\omega)$ between [0,1]. In this case, $T$ is very large and therefore we run 
   the simulations up to $t = 10\pi$, in steps of $dt = 2\pi / (2L^2)$. Surprisingly, the critical point 
   $R_{0c} = 0.908(5)$, the crossing probability $\approx 0.63$ and the set of critical exponents remain 
   unaltered within our numerical accuracy i.e., they do not depend on the actual number of distinct
   frequencies. 

      Here, we put forward an explanation for this frequency independence. Let $p(R)$ be the probability
   distribution of the radii of the discs which we argue to be independent of time using Eqn.\ \ref{EQN01}.
   Introducing a variable $Q = \omega t$ the joint distribution function $p(Q,R)$ can be expressed in terms 
   of the distribution functions of the two mutually independent variables $Q$ and $\phi$ as, $p(Q,R) = 
   p(\omega t)p(\phi)|J(Q,R)|$, where $J(Q,R)$ is the Jacobian of the transformation. Finally, the marginal 
   distribution of $R$ is calculated from $p(Q,R)$ and has the form
\begin{equation}
p(R) = 1 \Big/ \Big(\pi \sqrt{RR_0-R^2}\Big),
\label{EQN11}
\end{equation}   
   independent of the distribution of $p(\omega)$. This result can be compared for a system having a uniform
   distribution of disc radii between $[0,R_0]$, where the transition occurs at $R_{0c} = 0.925(5)$ \cite{Kundu}.
   Eqn.\ \ref {EQN11} has been verified numerically and the matching is very good (not shown here). Using 
   this equation one can calculate the probability that a bond is connected by the sum rule. Equating this 
   probability to 1/2, the random bond percolation threshold and neglecting local correlations one obtains
   an approximate estimate of $R_{0c} = 1$.

   \section{The second percolation transition}

      In this section we exhibit that a second percolation transition exists in terms of the passage time
   for information propagation. For this description we consider that an information propagates with
   infinite speed within a cluster of connected bonds i.e., spreads instantly to all sites of the cluster
   irrespective of the site of its introduction. This implies that for $R_0 > R_{0c}$ there exists a
   spanning cluster across the system through which an information can be transmitted at the same
   time instant from one side of the system to its opposite side. On the other hand for $R_0 < R_{0c}$ there
   are finite isolated clusters of connected bonds which dynamically change their shapes and sizes.

      Now we introduce the second mechanism for information propagation. We assume that the sites of an
   isolated informed cluster of connected bonds retain the information with themselves forever.
   In a latter time during the time evolution, this informed cluster may merge with another uninformed
   cluster and information would then propagate instantly to the sites of the new cluster. It is therefore
   apparent that if one waits long enough, may be several multiples of the time period $T$, it is likely
   that information would propagate through the system even when $R_0 < R_{0c}$. More elaborately, all
   sites at the top row of the square lattice are given some information at time $t = 0$. This
   information is instantly transmitted to all sites of all clusters that have at least one site on the
   top row. All these sites are now informed sites and they keep the information with them. Since time is
   increased in steps of $dt$, at the next time step the status of every bond is freshly determined
   and some new sites / clusters may get linked to these informed sites through a fresh set of connected bonds.
   Immediately, the information is transmitted again to all sites of all these clusters. In this way the
   information spreads to more and more sites of the entire lattice. Sometimes it may happen that the spreading
   process pauses for few time steps, though the status of different bonds are still changing. We assume 
   that the spreading process terminates permanently when the information reaches the bottom of the lattice. 
   The time required on average for this passage is denoted by $T_I(R_0,L)$. Since the average number of 
   connected bonds in the system decreases when the value of $R_0$ is decreased, this average information 
   passage time increases. Finally, $T_I(R_0,L)$ diverges when $R_0$ approaches $R_0^*$ from above. 
   Therefore, we recognize $R_0^*$ as the second critical point of percolation transition.
   
%--------------------------------------------------------------------------
\begin{figure}[t]
\begin{center}
\includegraphics[width=\linewidth]{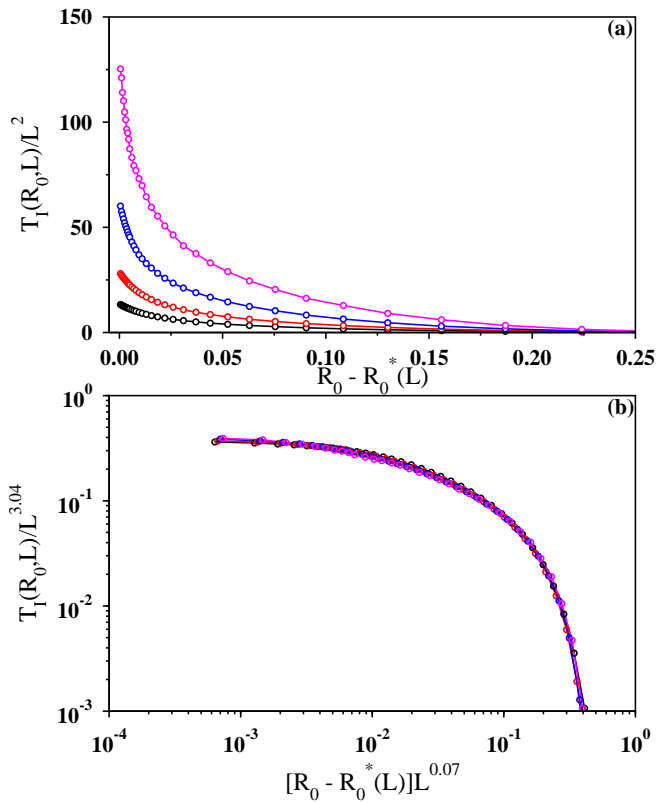}
\end{center}
\caption{(a) Average passage time for information propagation $T_I(R_0,L) / L^2$ has been plotted against the deviation 
   from the critical point $R_0 - R_0^*(L)$ for $L$ = 32 (black), 64 (red), 128 (blue) and 256 (magenta) (arranged from 
   bottom to top) using $\omega 
   = 1$ for all the discs. As $R_0 \to R_0^*(L)$, the time $T_I(R_0,L)$ diverges. (b) A scaling by $T_I(R_0,L)/L^{3.04}$ 
   against $(R_0 - R_0^*(L))L^{0.07}$ exhibits a good data collapse.}
\label{FIG07}
\end{figure}
%---------------------------------------------------------------------------

      In general for $R_0 > 1/2$, the live and dead status of all bonds of the lattice are determined. This
   gives a frozen configuration of live and dead bonds for every initial configuration of random phase angles.
   Only the live bonds can take part in the information propagation, and therefore, for a global passage of
   information across the system, it is necessary that the system must have a spanning cluster of live bonds.
   This leads us to identify the critical point $R_0^*$ as the configuration averaged minimum value of $R_0$ when a spanning
   cluster of live bonds appears in the system. Numerically, the precise value of $R_0^*$ has been estimated 
   using the bisection method. We started with a pair of values of $R_0$, namely $R_0^{hi}$ and $R_0^{low}$,
   corresponding to the globally connected and disconnected system respectively through the live bonds. 
   This interval is iteratively bisected till it becomes smaller than a pre-assigned tolerance value of $10^{-7}$.
   Averaging over a large number of independent configurations $R_0^*(L)$ for the system size $L$ is estimated. 
   The entire procedure is then repeated for several values of $L$ and extrapolated to $L \to \infty$ to 
   obtain $R_0^* = R_0^*(\infty)$. We find that the usual extrapolation method using $L^{-1/\nu}$ works very 
   well here as well with $\nu = 4/3$. Our best estimate for the critical point is $R_0^* = 0.5907(3)$.

      The average information propagation time $T_I(R_0,L)/L^2$ has been plotted against $R_0 - R_0^*(L)$ in Fig.\ \ref{FIG07}(a) 
   for four different system sizes using $\omega = 1$ for all the discs. It is observed that as $R_0$ approaches 
   $R_0^*$, the propagation time becomes increasingly larger. Further, for a specific value of $R_0$, the propagation 
   time increases with the system size. In Fig.\ \ref{FIG07}(b) the scaled plot of the same data has been exhibited. A 
   data collapse is obtained when $T_I(R_0,L)/L^{3.04}$ has been plotted against $(R_0 - R_0^*(L))L^{0.07}$. This is 
   consistent with the variation of the largest passage time which grows as $T_I(R_0^*,L) \sim L^{3.08}$.

      To characterize precisely the second transition point in terms of the live bonds, we have also estimated 
   the fractal dimension $d_f$ of the largest cluster of live bonds, the exponent $\gamma$ for the second 
   moment of the cluster size distribution at $R_0^*$, and the order parameter exponent $\beta$ around $R_0^*$. These
   exponents are very much consistent with the ordinary percolation exponents in two dimensions.

      An approximate estimate of the critical point $R_0^*$ can also be made neglecting the local correlations. 
   At the transition point, the density of live bonds $p_l(R_0^*)$ is equated to 1/2, the random bond percolation
   threshold on the square lattice. Using Eqn.\ \ref{EQN05} we obtain $R_0^* = 2/(2+\sqrt{2}) \approx 0.5858$, 
   which is very close to our numerically obtained value of $R_0^* = 0.5907(3)$.

%--------------------------------------------------------------------------
\begin{figure}[t]
\begin{center}
\includegraphics[width=\linewidth]{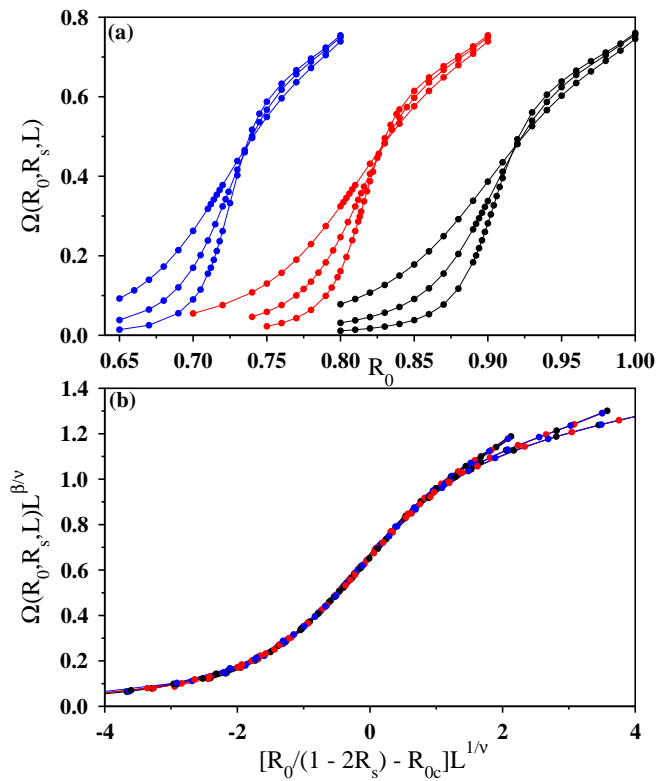}
\end{center}
\caption{For $R_s$ = 0.001 (black), 0.05 (red) and 0.1 (blue) (arranged from right to left), and for $L$ = 64, 128 and 256 for each $R_s$. (a) The 
   variation of the order parameter $\Omega(R_0,R_s,L)$ with $R_0$ has been shown using $\omega = 1$ for all the discs. 
   (b) The same data as in (a) has been scaled suitably. A scaling by $\Omega(R_0,R_s,L) L^{\beta/\nu}$ against 
   $[R_0 / (1-2R_s) - R_{0c}]L^{1/\nu}$ with $R_{0c} = 0.908(5)$, $1/\nu=0.75$ and $\beta/\nu = 0.112(5)$, exhibiting 
   a nice data collapse.} 
\label{FIG08}
\end{figure}
%--------------------------------------------------------------------------- 
   
\section{Generalized oscillating percolation}

      In this section we have generalized our model by introducing a shift parameter
   that enhances the disc radii by an amount $R_s$. Therefore, the radius of a disc 
   sinusoidally varies between $R_s$ and $R_0 + R_s$ as:
\begin{equation}
R(t) = R_s + (R_0/2)\{\sin(\omega t + \phi)+1\}.
\label{EQN12}
\end{equation}    
   For a specific value of $R_s$, it is now more likely that the radii of the ends discs of a bond would satisfy the
   Sum rule. Therefore the density of connected bonds at any given instant of time $t$ gets enhanced. As a consequence
   value of the critical amplitude $R_{0c}(R_s)$ decreases from its value $R_{0c}$ for $R_s = 0$.

      Variation of the order parameter $\Omega(R_0,R_s,L)$ has been studied against $R_0$ for three different 
   shifts $R_s$. For each $R_s$ three different system sizes $L$ have been exhibited in Fig.\ \ref{FIG08}(a). Using 
   the same data, in Fig.\ \ref{FIG08}(b) we show the scaling form
\begin{equation}
\Omega(R_0,R_s,L)L^{\beta/\nu} \sim {\cal F} \big[(R_0 / (1-2R_s) - R_{0c})L^{1/\nu}\big]
\label{EQN14} 
\end{equation}
   for the order parameter works very well with $R_{0c} = 0.908(5)$. The best data collapse is obtained using $1/\nu = 0.75$ 
   and $\beta/\nu = 0.112(5)$. Again, the finite-size scaling exponents closely match with the exponents of the 
   ordinary percolation in two dimensions.

   From Fig.\ \ref{FIG08}(b) equating $[R_0 / (1-2R_s) - R_{0c}]L^{1/\nu} = 0$ one gets
\begin{equation}
R_{0c}(R_s) = (1 - 2R_s)R_{0c}.
\label{EQN15}
\end{equation}
   Numerical values of $R_{0c}(R_s)$ are in very good agreement with those obtained from Eqn.\ \ref{EQN15}. 
   The shift $R_s$ effectively reduces the lattice constant by an amount $2R_s$. This explains the origin 
   of the factor $(1-2R_s)$ in Eqn.\ \ref{EQN15}.

\section{Summary}

      We have formulated a percolation model using a collection of pulsating discs keeping in mind the global 
   connectivity properties of the wireless sensor networks in the presence of temporal fluctuations of radio 
   transmission ranges.
   Every site of a regular lattice is occupied by a circular disc which pulsates sinusoidally within $[0,R_0]$. 
   The initial state is characterized by the random phase angles of the pulsating discs. Further, a lattice bond 
   is said to be connected as long as the pair of end discs overlap. The maximal radius $R_0$ acts as the control
   variable whose value is tuned continuously to change the fraction of the connected bonds in the system. The 
   first transition takes place at $R_{0c} = 0.908(5)$ when the giant cluster of connected bonds spans the entire
   system. It is imagined that the information passes through the spanning cluster instantly i.e., with infinite
   speed for all $R_0 > R_{0c}$. Moreover, the information can even transmit through the system when there are only
   isolated finite size clusters of connected bonds for $R_0 < R_{0c}$. This happens when informed clusters come
   in contact with the uninformed clusters and pass the information. Such transmission takes finite time to cover
   the system and it diverges when $R_0$ approaches $R_0^*$ from above, $R_0^*$ marks the second transition point.
   A consideration of the phase differences between the end discs of bonds allows one to classify all bonds in
   terms of dead and live. Dead bonds can never be connected, whereas the live bonds are connected at least once
   within one period. Interestingly, we could recognize $R_0^*$ to be the transition point when the global 
   connectivity through the spanning cluster of live bonds first appears in the system. Expectedly, both the transitions 
   exhibit the critical behavior of ordinary percolation transition since the interaction is short ranged.

      For the future investigations, one can generalize this model by placing the centers of the pulsating
   discs at random positions on a continuous plane by a Poisson process, like in continuum percolation 
   \cite{Meester,Stanley,Quintanilla}.

\section*{ACKNOWLEDGMENT}

      We thankfully acknowledge R. M. Ziff for his valuable comments and critical review of the manuscript.

\end {document}